\documentclass[10pt,showpacs,twocolumn,nofootinbib,aps,prd]{revtex4-1}
\usepackage{amsmath,amssymb,bm,mathrsfs}
\usepackage{mathtools,mathptmx,array,slashed}
\usepackage{graphicx,graphics,subfigure}
\usepackage{ctable,multirow}
\usepackage{tabularx}
\usepackage[colorlinks=true,linkcolor=blue,citecolor=blue,urlcolor=blue]{hyperref}

\DeclareSymbolFont{symbols} {OMS}{cmsy}{m}{n}

\def\be{\begin{equation}}
\def\ee{\end{equation}}
\def\bea{\begin{eqnarray}}
\def\eea{\end{eqnarray}}
\def\ba{\begin{aligned}}
\def\ea{\end{aligned}}
\def\nn{\nonumber}
\def\p{\partial}

\begin{document}


\title{Universal thermodynamic topological classes of rotating black holes}

\author{Xiao-Dan Zhu$^{1}$}

\author{Wentao Liu$^{2}$}

\author{Di Wu$^{1}$}
\email{Corresponding author: wdcwnu@163.com}

\affiliation{$^{1}$School of Physics and Astronomy, China West Normal University, Nanchong, Sichuan 637002, People's Republic of China \\
$^{2}$Department of Physics, Key Laboratory of Low Dimensional Quantum Structures and Quantum Control of Ministry of Education, and Synergetic Innovation Center for Quantum Effects and Applications, Hunan Normal University, Changsha, Hunan 410081, People's Republic of China}

\date{\today}

\begin{abstract}
In a recent study, Wei \textit{et al.} [\href{https://doi.org/10.1103/PhysRevD.110.L081501}
{Phys. Rev. D \textbf{110}, L081501 (2024)}] proposed a universal classification scheme that interprets black hole solutions, including the four-dimensional Schwarzschild, Reissner-Nordstr\"om (RN), Schwarzschild-AdS, and RN-AdS black hole solutions, as topological defects within the thermodynamic parameter space, and then divides black hole solutions into four distinct classes, denoted as $W^{1-}$, $W^{0+}$, $W^{0-}$, and $W^{1+}$, offering insights into deeper aspects of black hole thermodynamics and gravity. In this paper, we investigate the universal thermodynamic topological classification of the singly rotating Kerr black holes in all dimensions, as well as the four-dimensional Kerr-Newman black hole. We show that the innermost small black hole states of the $d \geq 6$ singly rotating Kerr black holes are thermodynamically unstable, while those of the four-dimensional Kerr-Newman black hole and the $d = 4, 5$ singly rotating Kerr black holes are thermodynamically stable. On the other hand, the outermost large black holes exhibit unstable behavior in all these cases. At the low-temperature limit, the $d \geq 6$ singly rotating Kerr black holes have one large thermodynamically unstable black hole state, while the four-dimensional Kerr-Newman black hole and the $d = 4, 5$ singly rotating Kerr black holes feature one large unstable branch and one small stable branch. Conversely, at the high-temperature limit, the $d \geq 6$ singly rotating Kerr black holes exhibit a small unstable black hole state, while the four-dimensional Kerr-Newman black hole and the $d = 4, 5$ singly rotating Kerr black holes have no black hole states at all. Consequently, we demonstrate that the $d \geq 6$ singly rotating Kerr black holes belong to the class $W^{1-}$, whereas the four-dimensional Kerr-Newman and $d = 4, 5$ singly rotating Kerr black holes belong to the class $W^{0+}$, thereby further support the conjecture proposed in [\href{https://doi.org/10.1103/PhysRevD.110.L081501}{Phys. Rev. D \textbf{110}, L081501 (2024)}].
\end{abstract}

\maketitle


\section{Introduction}
Black holes, among the most extraordinary and captivating objects in the universe, have long been the focus of intensive theoretical and observational research. On the observational side, recent breakthroughs include direct imaging of black hole shadows \cite{PRL115-211102,PRD89-124004,
CQG40-165007,2406.00579,2407.07416} by the Event Horizon Telescope (EHT) \cite{APJL875-L1,
APJL930-L12} and the detection of gravitational waves resulting from black hole mergers \cite{PRL116-061102,PRL116-241103}. On the theoretical side, recent advances in understanding the topology of black holes have provided novel insights into the nature of gravity, with significant contributions from studies on light rings \cite{PRL119-251102,PRL124-181101,PRD102-064039,
PRD103-104031,PRD104-044019, PRD105-024049,PRD105-064070,PRD108-104041,PRD109-064050,2408.05569}, timelike circular orbits \cite{PRD107-064006,JCAP0723049,2406.13270}, thermodynamic phase transitions \cite{PRD105-104003,PRD105-104053,PLB835-137591,PRD107-046013,PRD107-106009,
JHEP0623115, 2305.05595,2305.05916,2305.15674,2305.15910,2306.16117,PRD106-064059,PRD107-044026, PRD107-064015,2212.04341,2302.06201,2304.14988,2309.00224,2312.12784,2402.18791,2403.14730, 2404.02526,2407.09122,2407.20016,2408.03090,2408.03126,NPB1006-116653,2408.05870}, and special thermodynamic topological classifications \cite{PRL129-191101,PRD107-024024,PRD107-084002,
EPJC83-365,2306.02324,PRD108-084041,2402.00106,PLB856-138919,PDU46-101617,2409.11666,
PRD107-064023,JHEP0123102}. \footnote{The black hole solutions are classified into three categories based on different topological numbers. Please see more examples in Refs. \cite{PRD107-084053,2303.06814,2303.13105,2304.02889,2306.13286,2304.05695,2306.05692,2306.11212,
2307.12873,2309.14069,AP458-169486,2310.09602,2310.15182,2311.04050,2311.11606,2312.04325,
2312.06324,2312.13577,2312.12814,PS99-025003,2401.16756,AP463-169617,PDU44-101437,2403.14167,
2405.02328,2405.07525,2405.20022,2406.08793,AC48-100853,2407.05325,2408.08325,2409.04997,
EPJP139-806} for the latest developments.}

Quite recently, in Ref. \cite{2409.09333}, Wei \textit{et al.} further developed the method initially proposed in Ref. \cite{PRL129-191101} by using the generalized off-shell free energy to treat black hole solutions as topological thermodynamic defects. They analyzed the asymptotic behaviors of the constructed vector and, using four-dimensional Schwarzschild, Reissner-Nordstr\"om (RN), Schwarzschild-AdS, and RN-AdS black holes as examples, categorized black hole solutions into four distinct topological classes: $W^{1-}$, $W^{0+}$, $W^{0-}$, $W^{1+}$. \footnote{For the details of the four topological classes, please see Appendix \ref{Appa}.} This classification offers new insights into the fundamental nature of black hole thermodynamics and gravity. A brief introduction of this approach is presented below.

According to Ref. \cite{2409.09333}, one can first introduce the generalized off-shell Helmholtz free energy as \cite{PRD106-106015}
\be\label{FE}
\mathcal{F} = M -\frac{S}{\tau} \, ,
\ee
where $M$ and $S$ represent the mass and entropy of the black hole, respectively, and the additional variable $\tau$ can be interpreted as the inverse temperature of the cavity enclosing the black hole. The generalized Helmholtz free energy exhibits only on-shell properties and reduces to the standard Helmholtz free energy $F = M -TS$ \cite{PRD15-2752,PRD33-2092} of the black hole when $\tau = \beta = T^{-1}$. By including an additional parameter $\Theta\in(0,\pi)$, one can define a two-dimensional vector $\phi$ based on the gradient of $\mathcal{\hat{F}} = \mathcal{F} +1/\sin\Theta$, given by
\be\label{vector}
\phi = \left(\frac{\p\mathcal{\hat{F}}}{\p r_h}, ~\frac{\partial \mathcal{\hat{F}}}{\p\Theta} \right),
\ee
where $r_h$ is the event horizon radius of the black hole. A thorough analysis demonstrates that the black hole states precisely align with the zero points of the vector $\phi$. Then, by applying Duan's $\phi$-mapping topological current theory \cite{SS9-1072,NPB514-705,PRD61-045004}, a topological invariant, also referred to as the winding number $w$, can be attributed to each zero point or black hole state \cite{PRL129-191101}. Utilizing the above framework, the heat capacity of a black hole state can be either positive or negative, corresponding to winding numbers of $w = +1$ and $w = -1$, respectively. Here, a positive winding number indicates a locally stable black hole state, while a negative winding number signifies a locally unstable state. The topological number $W$, defined as
\be
W = \sum_{i=1}^{N}w_{i} \, ,
\ee
is obtained by summing all the winding numbers $w_i$ associated with the $i$th zero point of the field $\phi$, where $N$ is the total number of zero points. This topological number $W$ serves as a classification parameter for black hole systems, thereby providing a novel way to classify different black hole solutions. It is important to highlight that the local winding number $w_i$ serves as an effective tool for characterizing local thermodynamic stability. Thermodynamically stable black holes correspond to positive $w_i$ values, while unstable black holes correspond to negative values. Conversely, the global topological number $W$ reflects the difference between the quantities of stable and unstable black holes within a classical solution at a fixed temperature. Thus, the local winding number differentiates between stable and unstable phases of black holes at a given temperature and aids in classifying black hole solutions based on the global topological number. Furthermore, black holes within the same universal thermodynamic topological class exhibit similar thermodynamic properties, regardless of their geometric classification. This finding not only highlights the intrinsic connections between different black hole solutions but also provides a new perspective for studying the thermodynamic phase transitions and stability of black holes. Through this classification, we can better understand the mechanisms behind black hole behavior and their impact on thermodynamic stability and phase transitions. Since astronomical observations have largely focused on rotating black holes, exploring their universal thermodynamic topological classes is both essential and necessary. This provides the main motivation for this paper.

In this paper, we investigate the universal thermodynamic topological classification of the singly rotating Kerr black holes in arbitrary dimensions, and the four-dimensional Kerr-Newman black hole. We find that the $d \geq 6$ singly rotating Kerr black holes belong to the class $W^{1-}$, while the four-dimensional Kerr-Newman and $d = 4, 5$ singly rotating Kerr black holes belong to the class $W^{0+}$. Compared to our previous work \cite{PRD107-024024}, we provide new insights into the thermodynamic stability of the innermost small black hole state and the outermost large black hole state, as well as their thermodynamic properties in both the low-temperature and high-temperature limits. Understanding these aspects is crucial for revealing the complex behavior of rotating black holes, as it helps elucidate their phase transitions and stability conditions. Furthermore, we explore the systematic orderings of the black hole states corresponding to increasing horizon radius, which sheds light on the underlying mechanisms governing black hole thermodynamics. The remaining part of this paper is organized as follows.
In Sec. \ref{II}, we explore the universal thermodynamic topological class of the four-dimensional Kerr black hole and analyze its state systematic ordering and universal thermodynamic behavior.
In Secs. \ref{III} and \ref{IV}, this analysis is extended to $d$-dimensional singly rotating Kerr black holes and the four-dimensional Kerr-Newman black hole, respectively. Sec. \ref{V} provides a summary of our findings and outlooks for future work.

\section{Four-dimensional Kerr black hole}\label{II}
In this section, we will focus on the universal thermodynamic topological class of the four-dimensional Kerr black hole, whose metric in the asymptotically nonrotating frame
has the form \cite{CMP10-280}
\bea\label{Kerr}
ds^2 &=& -\frac{\Delta_r}{\Sigma}\big(dt -a\sin^2\theta\, d\varphi \big)^2
 +\frac{\Sigma}{\Delta_r}dr^2 +\Sigma\, d\theta^2 \nn \\
&& +\frac{\sin^2\theta}{\Sigma}\big[adt -\big(r^2 +a^2\big)d\varphi \big]^2 \,
\eea
where
\bea
\Delta_r = r^2 +a^2 -2mr \, , \quad \Sigma = r^2 +a^2\cos^2\theta \, , \nn
\eea
in which $m$ and $a$ are the mass and the rotation parameters of the black hole.

The thermodynamic quantities of the four-dimensional Kerr black hole can be derived using standard techniques, yielding the following elegant expressions \cite{CMP31-161,PRL30-71}:
\be\ba\label{ThermKerr}
&M = m \, , \quad J = ma \, , \quad \Omega = \frac{a}{r_h^2 +a^2} \, , \\
&S = \pi(r_h^2 +a^2) \, , \quad  T = \frac{r_h^2 -a^2}{4\pi r_h(r_h^2 +a^2)} \, ,
\ea\ee
where the event horizon is located at $r_h = m + \sqrt{m^2 - a^2}$. It is easy to find that the Hawking temperature of the four-dimensional Kerr black hole approaches zero both in the limit $r \to \infty$ and in the extremal case, where $r \to r_m = m = a$, with $r_m$ being the minimal horizon radius of the four-dimensional Kerr black hole. Consequently, the inverse temperature $\beta(r_h)$ exhibits
\be\label{c1}
\beta(r_m) = \infty \, , \qquad \beta(\infty) = \infty
\ee
at these asymptotic limits. The defect curve $\beta(r_h)$ must remain analytic over the range ($r_m$, $\infty$).

Using the definition of the generalized off-shell Helmholtz free energy (\ref{FE}), one can easily obtained
\be
\mathcal{F} = \frac{r_h^2 +a^2}{2r_h} -\frac{\pi(r_h^2 +a^2)}{\tau}
\ee
for the four-dimensional Kerr black hole. Thus,
\be
\mathcal{\hat{F}} = \mathcal{F} +\frac{1}{\sin\Theta} = \frac{r_h^2 +a^2}{2r_h} -\frac{\pi(r_h^2 +a^2)}{\tau} +\frac{1}{\sin\Theta} \, .
\ee
From Eq. (\ref{vector}), the components of the vector $\phi$ can be computed as
\bea
&&\phi^{r_h} = \frac{1}{2} -\frac{a^2}{2r_h^2} -\frac{2\pi r_h}{\tau} \, ,  \label{phir} \\
&&\phi^{\Theta} = -\cot\Theta\csc\Theta \, .
\eea
Therefore, the defect curve, which represents the set of the zero points of the vector in the parameter space, can be readily obtained by solving Eq. (\ref{phir}).

We now examine the asymptotic behavior of the vector $\phi$ at the boundary corresponding to Eq. (\ref{c1}).
This boundary can be described by the contour $C = I_1 \cup I_2 \cup I_3 \cup I_4$, where $I_1 = \{r_h = \infty, \Theta \in (0, \pi)\}$, $I_2 = \{r_h \in (\infty, r_m), \Theta = \pi\}$, $I_3 = \{r_h = r_m, \Theta \in (\pi, 0)\}$, and $I_4 = \{r_h \in (r_m, \infty), \Theta = 0\}$.
This contour covers all relevant parameter regions. The setup of $\phi$ ensures that it is orthogonal to $I_2$ and $I_4$ \cite{PRL129-191101}, so the key asymptotic behavior lies along $I_1$ and $I_3$. As $r_h$ approaches $r_m$ and $\infty$, the direction of $\phi$ shifts leftward, with an inclination that depends on the value of $\phi^\Theta$. In Table \ref{TableI}, the direction pairs on the segments $I_1$ and $I_3$, along with the topological number $W$, are listed for the four-dimensional Kerr black hole.

\begin{table}[t]
\caption{The direction indicated by the arrows of $\phi^{r_{h}}$ and the corresponding topological number for the four-dimensional Kerr black hole.}
\begin{tabular}{c|cccc|c}
\hline
Black hole solutions & $I_1$ & $I_2$ & $I_3$ & $I_4$ & $W$ \\ \hline
$d=4$ Kerr black hole & $\leftarrow$  & $\uparrow$ & $\leftarrow$ & $\downarrow$ & 0 \\
\hline
\end{tabular}
\label{TableI}
\end{table}

For the four-dimensional Kerr black hole, the topological number is $W = 0$. A generate point is found at $\beta_c/r_0 = 6\sqrt{3}\pi a$ \cite{PRD107-024024}, where $r_0$ is an arbitrary length scale determined by the size of a surrounding cavity. Below this point, no black hole states are present, indicating a trivial topology. At larger values of $\beta$, two black hole states--small and large--appear, with the former being thermodynamically stable and the latter unstable. The topological number always equals $0$, regardless of the values of $\tau$ and the rotation parameter $a$.

\begin{figure}[t]
\centering
\includegraphics[width=0.25\textwidth]{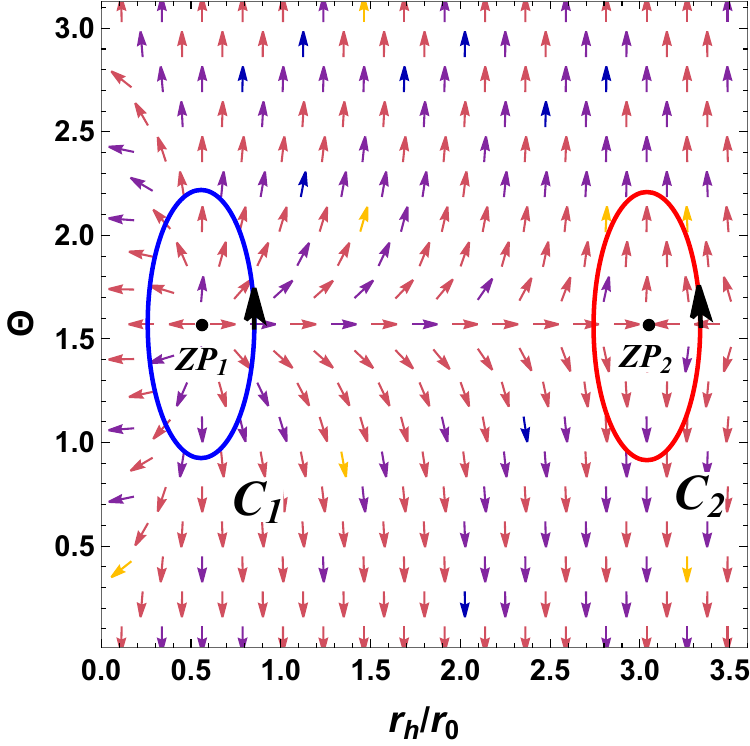}
\caption{The arrows represent the unit vector field on a portion of the $r_h-\Theta$ plane for the four-dimensional Kerr black hole with $\beta/r_0 = 40$ and $a/r_0 = 0.5$. The zero points (ZPs) marked with black dots are at $(r_h/r_0, \Theta) = (0.55, \pi/2)$, and $(3.10, \pi/2)$, for ZP$_1$, and ZP$_2$, respectively. The blue contours $C_i$ are closed loops surrounding the zero points.
\label{Fig1}}
\end{figure}

\begin{figure}[t]
\centering
\includegraphics[width=0.25\textwidth]{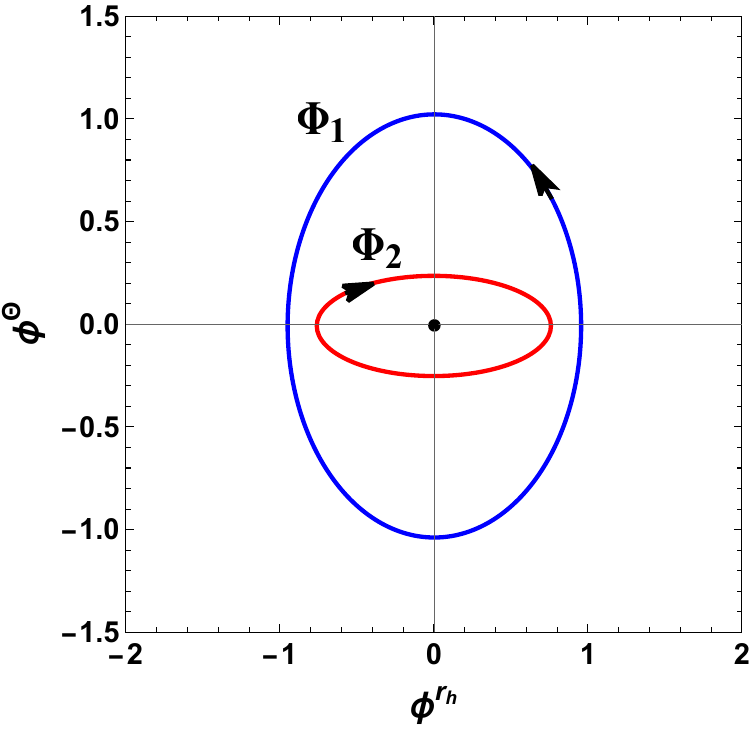}
\caption{Contours $\phi_i$ that represent changes in the components of the vector $\phi$ for the four-dimensional Kerr black hole are illustrated as $C_i$ ($i = 1,2$) in Fig. \ref{Fig1}. The origin denotes the zero points of $\phi$. The black arrows indicate the direction of rotation of the vector $\phi$ as it traverses each contour in Fig. \ref{Fig1}. Along contour $C_1$, the modifications in the components of $\phi$ cause the closed loop $\Phi_1$ to follow a counterclockwise direction, resulting in a winding number of $+1$. Conversely, for contour $C_2$, the behavior is reversed, yielding a winding number of $-1$.
\label{Fig2}}
\end{figure}

Consider the variations in the components $(\phi^{r_h},\phi^{\Theta})$ of the vector $\phi$ for the four-dimensional Kerr black hole case along each contour depicted in Fig. \ref{Fig1}, as illustrated in Fig. \ref{Fig2}. The zero points of $\phi$ are located at the origin, and tracking the changes of $\phi$ along each $C_i$ generates a closed loop $\Phi_i$ in the corresponding vector space. In Fig. \ref{Fig2}, the direction of each contour reflects the rotation of $\phi$ as the associated contour in Fig. \ref{Fig1} is traversed counterclockwise. Contours with positive winding numbers in Fig. \ref{Fig1} are mapped to counterclockwise loops in Fig. \ref{Fig2}, while those with negative winding numbers are mapped to clockwise loops. In the case of the four-dimensional Kerr black hole, the winding numbers of the first and second zero points are $+1$ and $-1$, respectively.

Then, we analyze the systematic ordering for the four-dimensional Kerr black hole. There is at least one black hole state with positive heat capacity and a winding number of $+1$, and one black hole state with negative heat capacity and a winding number of $-1$. If additional black hole states are present, they must emerge in pairs. As the signs of the heat capacities alternate with increasing $r_h$, the smallest state corresponds to a thermodynamically stable black hole, while the largest state corresponds to a thermodynamically unstable one. The winding numbers associated with the zero points follow the sequence $[+, (-, +), ..., -]$, where the ellipsis represents pairs of $(+, -)$ winding numbers. For simplicity, the four-dimensional Kerr black hole case can be labeled as $[+, -]$ based on the signs of the innermost and outermost winding numbers.

In the following, we turn to discuss the universal thermodynamic behavior of the four-dimensional Kerr black hole. In the low-temperature limit, $\beta \to \infty$, the system features an unstable large black hole and a stable small black hole. At the high-temperature limit, $\beta \to 0$, no black hole state is present.

To summarize, based on the thermodynamic topological classification approach proposed in Ref. \cite{2409.09333}, the four-dimensional Kerr black hole falls under the class $W^{0+}$.

\section{Singly rotating Kerr black holes in arbitrary dimensions}\label{III}
In this section, we will extend the above discussion to rotating black holes in higher dimensions, specifically considering singly rotating Kerr black holes in $d$ dimensions. The metric for these black holes in arbitrary dimensions is expressed as \cite{AP172-304,PRD93-084015}
\bea\label{hdKerr}
ds^2 &=& -\frac{\Delta_r}{\Sigma}\left(dt -a\sin^2\theta d\phi \right)^2
 +\frac{\Sigma}{\Delta_r}dr^2 +\Sigma d\theta^2 \nn \\
&& +\frac{\sin^2\theta}{\Sigma}\left[adt -(r^2 +a^2)d\phi \right]^2
 +r^2\cos^2\theta d\Omega_{d-4}^2 \, , \qquad
\eea
where $d\Omega_{d}$ represents the line element of a $d$-dimensional unit sphere, and
\bea
\Delta_r = r^2 +a^2 -2mr^{5-d} \, , \quad \Sigma = r^2 +a^2\cos^2\theta \, . \nn
\eea

The thermodynamic quantities are \cite{PRD93-084015}
\be\ba\label{ThermhdKerr}
&M = \frac{d-2}{8\pi}\omega_{d-2}m \, , \quad J = \frac{\omega_{d-2}}{4\pi}ma \, ,  \\
&\Omega = \frac{a}{r_h^2 +a^2} \, , \quad
 S = \frac{\omega_{d-2}}{4}(r_h^2 +a^2)r_h^{d-4} \, , \\
&T = \frac{r_h}{2\pi}\left(\frac{1}{r_h^2 +a^2}
 +\frac{d-3}{2r_h^2} \right) -\frac{1}{2\pi r_h} \, ,
\ea\ee
where $\omega_{d-2} = 2\pi^{(d-1)/2}/\Gamma[(d-1)/2]$, and the event horizon radius $r_h$ of the black hole is determined by the largest root of the equation: $\Delta_{r_h} = 0$.

Next, we analyze the asymptotic behavior of the inverse temperature $\beta(r_h)$ for singly rotating Kerr black holes in different dimensions. For the case of the five-dimensional singly rotating Kerr black hole, its Hawking temperature approaches zero in two situations: one in the limit $r \to \infty$, and the other in the extremal case $r \to r_m = \sqrt{2m - a^2}$. However, for the cases of $d \geq 6$ singly rotating Kerr black holes, the Hawking temperature approaches zero in the limit $r \to \infty$, but it diverges as $r \to 0$. Thus, it yields two possibilities:
\bea
&&d = 5~~case: \quad \beta(r_m) = \infty \, , \quad \beta(\infty) = \infty \, , \label{c2} \\
&&d \geq 6~~cases: \quad \beta(r_m) = 0 \, , \quad \beta(\infty) = \infty \label{c3}
\eea
for the asymptotic behaviour of the inverse temperature $\beta(r_h)$.

From Eq. (\ref{ThermhdKerr}), one can calculate the generalized free energy as
\bea
\mathcal{F} &=& M -\frac{S}{\tau} \nn \\
&=& \frac{\omega_{d-2}(d-2)(r_h^2 +a^2)}{16\pi r_h^{5-d}}
 -\frac{\omega_{d-2}(r_h^2 +a^2)r_h^{d-4}}{4\tau} \, . \quad
\eea
Therefore the components of the vector $\phi$ can be computed as
\bea
\phi^{r_h} &=& \frac{\omega_{d-2}r_h^{d-6}}{16\pi\tau}\Big\{(d-2)\big[(d-3)\tau
 -4\pi r_h \big]r_h^2 \nn \\
&& +a^2\big[\tau(d-2)(d-5)-4(d-4)\pi r_h \big] \Big\}  \, , \\
\phi^{\Theta} &=& -\cot\Theta\csc\Theta \, .
\eea

\begin{table}[t]
\caption{The direction indicated by the arrows of $\phi^{r_{h}}$ and the corresponding topological number for the $d\geq 5$ singly rotating Kerr black holes.}
\begin{tabular}{c|cccc|c}
\hline
Black hole solutions & $I_1$ & $I_2$ & $I_3$ & $I_4$ & $W$ \\ \hline
$d=5$ singly rotating Kerr black hole & $\leftarrow$  & $\uparrow$ & $\leftarrow$ & $\downarrow$ & 0 \\
$d\geq 6$ singly rotating Kerr black hole & $\leftarrow$  & $\uparrow$ & $\rightarrow$ & $\downarrow$ & -1 \\
\hline
\end{tabular}
\label{TableII}
\end{table}

We now analyze the asymptotic behavior of the vector $\phi$ at the boundary corresponding to Eqs. (\ref{c2}) and (\ref{c3}), which is also described by the contour $C = I_1 \cup I_2 \cup I_3 \cup I_4$ that covers all relevant parameter regions. In Table \ref{TableII}, the direction pairs on segments $I_1$ and $I_3$, as well as the topological number $W$, are provided for the $d\geq 5$ singly rotating Kerr black holes.

Based on the conclusions in our previous work \cite{PRD107-024024}, the topological numbers of the five-dimensional singly rotating Kerr black holes differ from those of the $d \geq 6$ singly rotating Kerr black holes. Therefore, we will next discuss the universal thermodynamic topological classification and related properties of the five-dimensional singly rotating Kerr black holes and the $d \geq 6$ singly rotating Kerr black holes separately.

\subsection{$d = 5$ case}

In this subsection, we focus on the case of the five-dimensional singly rotating Kerr black hole.
For the $ d = 5 $ singly rotating Kerr black hole, the topological number is $ W = 0 $. A critical generation point occurs at $\beta_c/r_0 = 4\sqrt{3}\pi a/3$ \cite{PRD107-024024}, below which no black hole states exist, indicating trivial topology. At higher values of $ \beta $, two black hole states, small and large, appear. The small black hole is thermodynamically stable, while the large one is unstable.

\begin{figure}[t]
\centering
\includegraphics[width=0.25\textwidth]{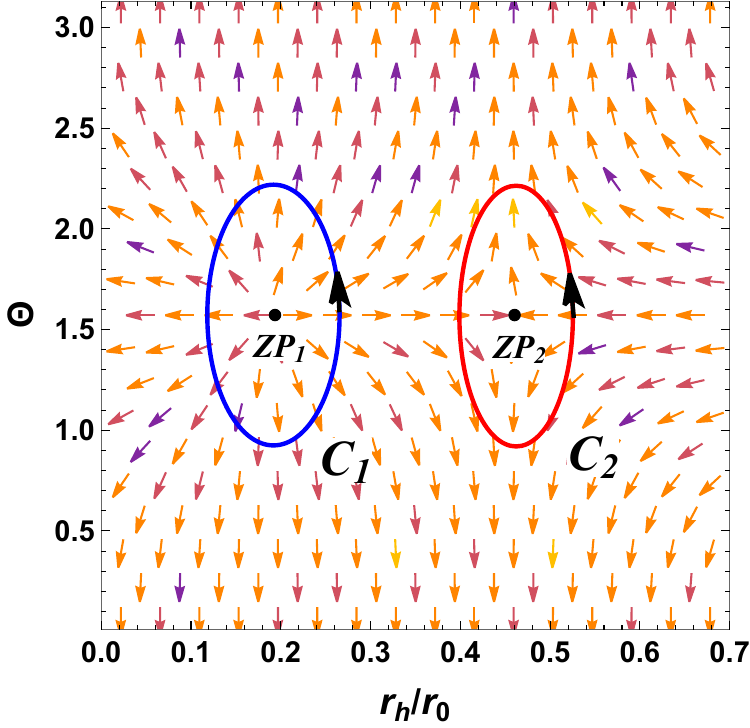}
\caption{The arrows represent the unit vector field on a portion of the $r_h-\Theta$ plane for the five-dimensional singly rotating Kerr black hole with $\beta/r_0 = 4$ and $a/r_0 = 0.5$. The zero points (ZPs) marked with black dots are at $(r_h/r_0, \Theta) = (0.19, \pi/2)$, and $(0.45, \pi/2)$, for ZP$_1$, and ZP$_2$, respectively. The blue contours $C_i$ are closed loops surrounding the zero points.
\label{Fig3}}
\end{figure}

\begin{figure}[t]
\centering
\includegraphics[width=0.25\textwidth]{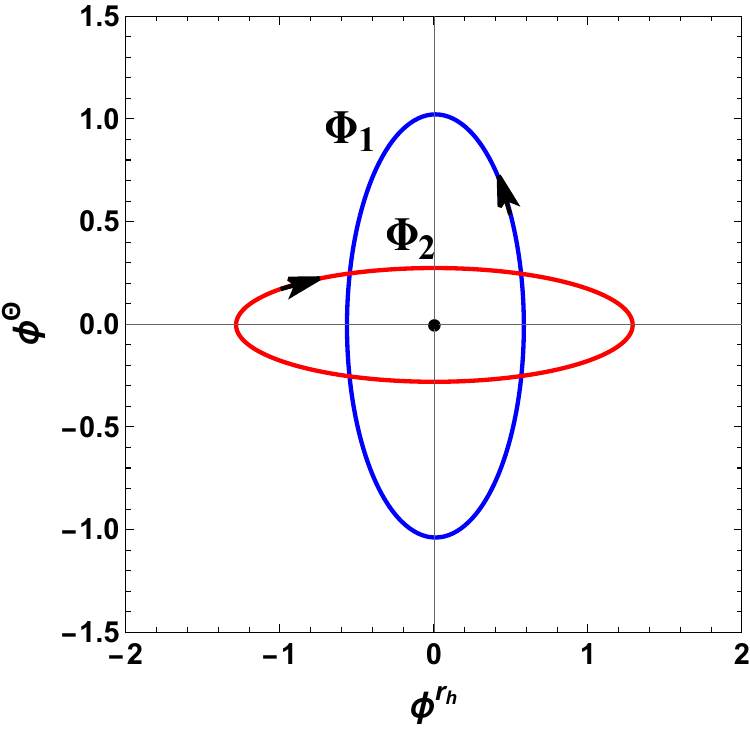}
\caption{Contours $\phi_i$, representing changes in the components of the vector $\phi$ for the five-dimensional singly rotating Kerr black hole, are shown as $C_i$ ($i = 1, 2$) in Fig. \ref{Fig3}. The origin denotes the zero points of $\phi$. Black arrows indicate the direction of rotation of the vector $\phi$ around each contour in Fig. \ref{Fig3}. For contour $C_1$, the changes in the components of $\phi$ cause the closed loop $\Phi_1$ to move counterclockwise, resulting in a winding number of $+1$. In contrast, for contour $C_2$, the movement is clockwise, giving a winding number of $-1$.
\label{Fig4}}
\end{figure}

Examine the behavior of the components $(\phi^{r_h},\phi^{\Theta})$ of the vector $\phi$ for the five-dimensional singly rotating Kerr black hole along each contour shown in Fig. \ref{Fig3}. This behavior is depicted in Fig. \ref{Fig4}. For this five-dimensional case, the winding numbers of the first and second zero points are $+1$ and $-1$, respectively, which are the same as the winding numbers for the four-dimensional Kerr black hole in the previous section.

We now examine the systematic ordering for the five-dimensional singly rotating Kerr black hole.
It is observed that there exists at least one black hole state with positive heat capacity and a winding number of $+1$, and one black hole state with negative heat capacity and a winding number of $-1$. For clarity, the five-dimensional singly rotating Kerr black hole can be classified as $[+,-]$ according to the signs of the innermost and outermost winding numbers.

We next address the universal thermodynamic behavior of the five-dimensional singly rotating Kerr black hole. In the low temperature limit, $\beta\to\infty$, the configuration features an unstable large black hole and a stable small black hole. In contrast, at high temperatures ($\beta\to 0$), no black hole states are observed.

In summary, according to the thermodynamic topological classification method outlined in Ref. \cite{2409.09333}, the five-dimensional singly rotating Kerr black hole is categorized as $W^{0+}$.

\subsection{$d\geq 6$ cases}

In this subsection, we turn to focus on the cases of the $d\geq 6$ singly rotating Kerr black hole. For these black holes, given a $\tau$, there exists only one black hole state characterized by a negative heat capacity. This state has a local winding number of $-1$, which is consistent with its global topological number of $-1$.

\begin{figure}[t]
\centering
\includegraphics[width=0.25\textwidth]{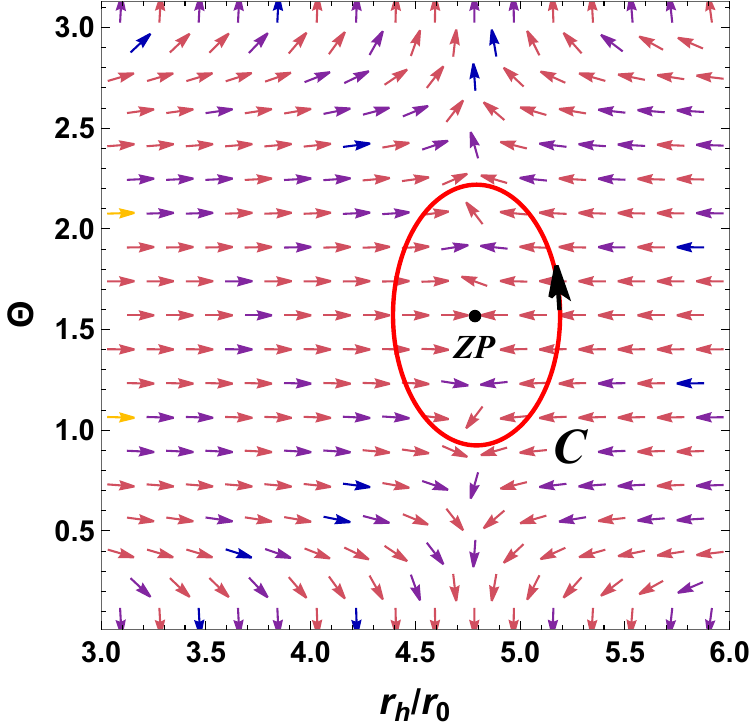}
\caption{The arrows represent the unit vector field on a portion of the $r_h-\Theta$ plane for the six-dimensional singly rotating Kerr black hole with $\beta/r_0 = 20$ and $a/r_0 = 0.5$. The zero point (ZP) marked with black dot is at $(r_h/r_0, \Theta) = (4.77, \pi/2)$. The blue contour $C$ is closed loop surrounding the zero point.
\label{Fig5}}
\end{figure}

\begin{figure}[t]
\centering
\includegraphics[width=0.25\textwidth]{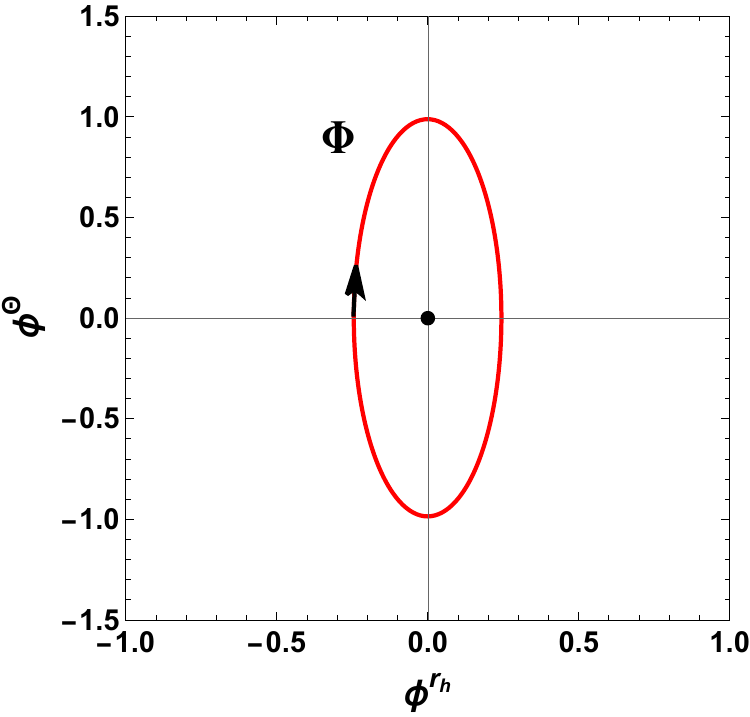}
\caption{Contours labeled $\phi_i$, representing the variations in the components of the vector $\phi$ for the six-dimensional singly rotating Kerr black hole, are denoted as $C$ in Fig. \ref{Fig5}. The origin marks the zero point of $\phi$. Black arrows illustrate the direction of the vector $\phi$'s rotation around each contour in Fig. \ref{Fig5}. For contour $C$, the changes in $\phi$ cause the closed loop $\Phi$ to trace a clockwise path, corresponding to a winding number of $-1$.
\label{Fig6}}
\end{figure}

As a specific example, we explore the behavior of the components $(\phi^{r_h},\phi^{\Theta})$ of the vector $\phi$ for the six-dimensional singly rotating Kerr black hole, analyzing each contour shown in Fig. \ref{Fig5} and illustrated in Fig. \ref{Fig6}. It is worth noting that the behavior of the components $(\phi^{r_h},\phi^{\Theta})$ of the vector $\phi$ for the seven-, eight-, and nine-dimensional singly rotating Kerr black holes is similar to that of the six-dimensional singly rotating Kerr black hole. Thus, we will not plot them again here. Therefore, we can conclude that for the $d\geq 6$ singly rotating Kerr black hole, the winding number of the zero point is always $-1$.

Now, we analyze the systematic ordering for the $d\geq 6$ singly rotating Kerr black holes. It is easy to observe that there is at least one black hole state that possesses a negative heat capacity and a winding number of $-1$. If additional black hole states exist, they must appear in pairs. The heat capacities alternate in sign with increasing $r_h$, with the smallest and largest states corresponding to thermodynamically unstable black holes. The winding numbers at the zero points follow the sequence $[-, (+, -), \ldots, (+,-)]$. For simplicity, the singly rotating Kerr black hole in dimensions $d \geq 6$ can be characterized by $[-, -]$, indicating the signs of the innermost and outermost winding numbers.

Then, we investigate the universal thermodynamic behavior of the $d\geq 6$ singly rotating Kerr black holes. In the low-temperature limit, where $\beta \to \infty$, the system features a large black hole that is thermodynamically unstable. Conversely, in the high-temperature limit, where $\beta \to 0$, the system displays an unstable small black hole state.

To sum up, based on the thermodynamic topological classification approach detailed in Ref. \cite{2409.09333}, the singly rotating Kerr black hole in $d \geq 6$ dimensions is classified as $W^{1-}$.

\section{Four-dimensional Kerr-Newman black hole}\label{IV}

Finally, in this section, we investigate the universal thermodynamic topological class of the four-dimensional Kerr-Newman black hole \cite{JMP6-915,JMP6-918}. Its metric and Abelian gauge potential are given by
\bea
ds^2 &=& -\frac{\Delta_r}{\Sigma}\left(dt -a\sin^2\theta d\phi \right)^2
 +\frac{\Sigma}{\Delta_r}dr^2 +\Sigma d\theta^2 \nn \\
&&+\frac{\sin^2\theta}{\Sigma}\left[adt -(r^2 +a^2)d\phi \right]^2 \, , \\
A &=& \frac{qr}{\Sigma}(dt -a\sin^2\theta d\phi) \, ,
\eea
where
\bea
\Delta_r = r^2 +a^2 -2mr +q^2 \, , \quad \Sigma = r^2 +a^2\cos^2\theta \, . \nn
\eea
Here, $m$ is the mass parameter. The parameters $a$ and $q$ represent the rotation and electric charge, respectively.

The thermodynamic quantities are expressed as follows \cite{CMP31-161,PRL30-71}:
\be\ba\label{ThermKN}
&M = m \, , \qquad J = ma \, , \qquad \Omega = \frac{a}{r_h^2 +a^2} \, , \\
&Q = q \, , \qquad \quad  \Phi = \frac{qr}{r_h^2 +a^2} \, , \\
&S = \pi(r_h^2 +a^2) \, , \qquad T = \frac{r_h^2 -a^2 -q^2}{4\pi r_h(r_h^2 +a^2)} \, ,
\ea\ee
where $r_h = m +\sqrt{m^2 -a^2 -q^2}$ is the location of the event horizon. The Hawking temperature of the four-dimensional Kerr-Newman black hole vanishes in both the limit $r \to \infty$ and the extremal horizon limit, where $r_m = m = \sqrt{a^2 + q^2}$.

As a result, the inverse temperature $\beta(r_h)$ reaches
\be\label{c4}
\beta(r_m) = \infty \, , \qquad \beta(\infty) = \infty
\ee
at these asymptotic boundaries.

Substituting $m = (r_h^2 +a^2 +q^2)/(2r_h)$  into the definition of the generalized Helmholtz free energy (\ref{FE}), one can arrive at
\be
\mathcal{F} = \frac{r_h^2 +a^2 +q^2}{2r_h} -\frac{\pi(r_h^2 +a^2)}{\tau} \, ,
\ee
and
\be
\mathcal{\hat{F}} = \frac{r_h^2 +a^2 +q^2}{2r_h} -\frac{\pi(r_h^2 +a^2)}{\tau} +\frac{1}{\sin\Theta} \, .
\ee
Thus, the components of the vector $\phi$ are
\bea
&&\phi^{r_h} = 1 -\frac{r_h^2 +a^2 +q^2}{2r_h^2} -\frac{2\pi r_h}{\tau} \, , \\
&&\phi^{\Theta} = -\cot\Theta\csc\Theta \, .
\eea

\begin{table}[b]
\caption{The direction indicated by the arrows of $\phi^{r_{h}}$ and the corresponding topological number for the four-dimensional Kerr-Newman black hole.}
\begin{tabular}{c|cccc|c}
\hline
Black hole solutions & $I_1$ & $I_2$ & $I_3$ & $I_4$ & $W$ \\ \hline
$d=4$ Kerr-Newman black hole & $\leftarrow$  & $\uparrow$ & $\leftarrow$ & $\downarrow$ & 0 \\
\hline
\end{tabular}
\label{TableIII}
\end{table}

\begin{figure}[t]
\centering
\includegraphics[width=0.25\textwidth]{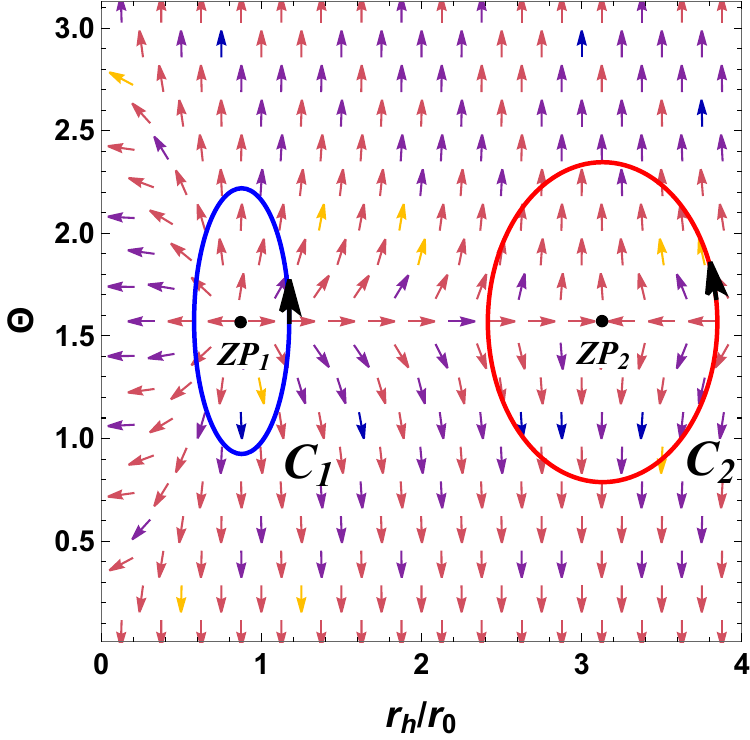}
\caption{The arrows represent the unit vector field on a portion of the $r_h-\Theta$ plane for the four-dimensional Kerr-Newman black hole with $\beta/r_0 = 40$, $a/r_0 = 0.5$ and $q/r_0 = 0.5$. The zero points (ZPs) marked with black dots are at $(r_h/r_0, \Theta) = (0.82, \pi/2)$, and $(3.01, \pi/2)$, for ZP$_1$, and ZP$_2$, respectively. The blue contours $C_i$ are closed loops surrounding the zero points.
\label{Fig7}}
\end{figure}

\begin{figure}[h]
\centering
\includegraphics[width=0.25\textwidth]{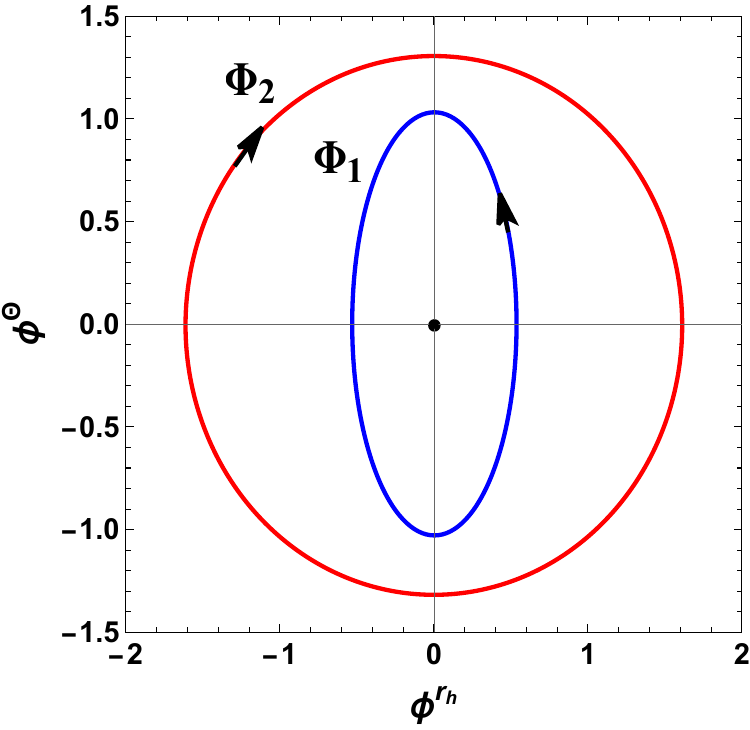}
\caption{Contours $\phi_i$, representing changes in the vector components for the four-dimensional Kerr-Newman black hole, are illustrated in Fig. \ref{Fig7} as $C_i$ ($i = 1, 2$). The origin marks the zero points of $\phi$. Black arrows indicate the rotation of the vector $\phi$ along each contour. For $C_1$, the changes in $\phi$ lead to a counterclockwise rotation of the closed loop $\Phi_1$, producing a winding number of $+1$. In contrast, for $C_2$, the loop rotates clockwise, giving a winding number of $-1$.
\label{Fig8}}
\end{figure}

The asymptotic behavior of the vector $\phi$ at the boundary associated with Eq. (\ref{c4}) is examined, with the contour $C = I_1 \cup I_2 \cup I_3 \cup I_4$ describing this behavior and encompassing all relevant parameter regions. Table \ref{TableIII} presents the direction pairs for segments $I_1$ and $I_3$, as well as the topological number $W$ for the four-dimensional Kerr-Newman black hole.

For the four-dimensional Kerr-Newman black hole, the topological number is $W = 0$.A generation point is found at $\beta_c/r_0 = 6\pi\sqrt{3(a^2 +q^2)}$ \cite{PRD107-024024}. Below this point, no black hole states exist, reflecting a trivial topology. At higher values of $\beta$, two black hole states emerge: a small, thermodynamically stable state and a large, unstable one. The topological number remains zero regardless of the values of $\tau$, the rotation parameter $a$, or the electric charge parameter $q$.

The behavior of the vector components $(\phi^{r_h},\phi^{\Theta})$ for the four-dimensional Kerr-Newman black hole is examined along each contour in Fig. \ref{Fig7}. This behavior is illustrated in Fig. \ref{Fig8}. In this case, the winding numbers at the first and second zero points are $+1$ and $-1$, respectively, which match those found for the four-dimensional Kerr black hole in Sec. \ref{II}.

We now investigate the systematic ordering for the four-dimensional Kerr-Newman black hole. It is demonstrated that there is always at least one black hole state with a positive heat capacity and a winding number of $+1$. Additionally, there is at least one state with a negative heat capacity and a winding number of $-1$. Therefore, the four-dimensional black hole is classified as $[+,-]$ based on the signs of its innermost and outermost winding numbers.

Next, we consider the universal thermodynamic behavior of the four-dimensional Kerr-Newman black hole. In the low-temperature limit, $\beta \to \infty$, the configuration consists of a stable small black hole and an unstable large black hole, respectively. At high temperatures ($\beta \to 0$), the black hole states disappear entirely.

In conclusion, following the thermodynamic topological classification method described in Ref. \cite{2409.09333}, the four-dimensional Kerr-Newman black hole is classified as $W^{0+}$.

\section{Conclusions and outlooks}\label{V}
Our results found in the present paper are now summarized in the following Table \ref{TableIV}.

In this paper, we explore the universal thermodynamic topological classification of singly rotating Kerr black holes in arbitrary dimensions, as well as the four-dimensional Kerr-Newman black hole. We find that the innermost small black hole states of the singly rotating Kerr black holes with $d \geq 6$ are thermodynamically unstable. In contrast, the corresponding states of the four-dimensional Kerr-Newman black hole and the $d = 4, 5$ singly rotating Kerr black holes  are thermodynamically stable. Additionally, the outermost large black holes are unstable in all cases.

At low temperatures, the $d \geq 6$ singly rotating Kerr black holes feature a single large, thermodynamically unstable black hole. In contrast, the four-dimensional Kerr-Newman black hole and the $d = 4, 5$ singly rotating Kerr black holes exhibit one large unstable branch and one small stable branch. At high temperatures, the $d \geq 6$ singly rotating Kerr black holes show a small, unstable black hole state, while the four-dimensional Kerr-Newman black hole and the $d = 4, 5$ singly rotating Kerr black holes have no black hole states.

As a result, we conclude that the singly rotating Kerr black holes with $d \geq 6$ are classified as $W^{1-}$, while the four-dimensional Kerr-Newman black hole and the $d = 4, 5$ singly rotating Kerr black holes with are classified as $W^{0+}$. This finding also supports the conjecture proposed in Ref. \cite{2409.09333}, which suggests that black hole solutions should be categorized into four distinct classes based on their universal thermodynamic topological properties.

A particularly intriguing issue is to delve deeper into the universal thermodynamic topological properties of black holes with unusual horizon topologies. These include planar \cite{PRD54-4891}, toroidal \cite{PRD56-3600}, hyperbolic \cite{PRD92-044058}, and ultraspinning black holes \cite{PRD89-084007,PRL115-031101,JHEP0114127,PRD103-104020,PRD101-024057,PRD102-044007,PRD103-044014,
JHEP1121031,PRD95-046002,JHEP0118042}, as well as NUT-charged spacetimes \cite{PRD100-101501,
PRD105-124013,2209.01757,2210.17504,2306.00062} and type-D NUT C-metric black holes
\cite{2409.06733}, etc. We believe that this could help elucidate the connections between horizon topology and thermodynamic topology.

\acknowledgments

We are greatly indebted to the anonymous referee for the constructive comments to improve the presentation of this work. This work is supported by the National Natural Science Foundation of China (NSFC) under Grants No. 12205243, No. 12375053, by the Sichuan Science and Technology Program under Grant No. 2023NSFSC1347, by the Doctoral Research Initiation Project of China West Normal University under Grant No. 21E028, and by the Postgraduate Scientific Research Innovation Project of Hunan Province under Grant No. CX20240531.

\appendix
\section{Four thermodynamic topological classes}\label{Appa}

According to Ref. \cite{2409.09333}, the four topological classes of black hole thermodynamics are:
\be
W^{1-} \, , \quad W^{0+} \, , \quad W^{0-} \, , \quad W^{1+} \, ,
\ee
These classes correspond to distinct asymptotic behaviors of the inverse temperature $\beta(r_h)$:
\bea
&W^{1-}&:~  \beta(r_{m})=0\, ,\quad\;\; \beta(\infty)=\infty \, ,  \\
&W^{0+}&:~  \beta(r_{m})=\infty \, ,\quad \beta(\infty)=\infty \, ,   \\
&W^{0-}&:~  \beta(r_{m})=0\, ,\quad\;\; \beta(\infty)=0 \, , \label{b0-} \\
&W^{1+}&:~  \beta(r_{m})=\infty \, ,\quad \beta(\infty)=0 \, , \label{b1+}
\eea
where $r_{m}$ represents the minimal radius of the black hole event horizon, which may or may not be zero. For instance, in the extremal case of a Reissner-Nordstr\"om (RN) black hole with a fixed charge $Q$, we have $r_{m} = M = Q = r_e$. In contrast, a Schwarzschild black hole has $r_{m}$ equal to zero. Table \ref{TableV} summarizes the properties of these four topological classes, thereby enhancing our understanding of their distinct characteristics.

\onecolumngrid
\begin{table*}[ht]
\centering
\caption{The universal thermodynamic topological classifications of the rotating black holes and their thermodynamical properties. Here and hereafter, DP, AP, and GP represent the degenerate point, annihilation point, and generation point, respectively.}
\begin{tabular}{c|c|c|c|c|c|c|c}\hline\hline
BH solutions & Innermost & Outermost & Low $T$ & High $T$ & DP & W & Classes \\ \hline
$d = 4$ Kerr BH & stable & unstable & unstable large~$+$~stable small & no & one more GP & 0 & $W^{0+}$ \\
$d = 4$ Kerr-Newman BH & stable & unstable & unstable large~$+$~stable small & no & one more GP & 0 & $W^{0+}$ \\
$d = 5$ singly rotating Kerr BH & stable & unstable & unstable large~$+$~stable small & no & one more GP & 0 & $W^{0+}$ \\
$d \geq 6$ singly rotating Kerr BH & unstable & unstable & unstable large & unstable small & in pairs & -1 & $W^{1-}$ \\
\hline\hline
\end{tabular}
\label{TableIV}
\end{table*}
\begin{table*}[t]
\centering
\caption{Thermodynamical properties of the black hole states for the four topological classes of $W^{1-}$, $W^{0+}$, $W^{0-}$, and $W^{1+}$, respectively.}\label{TableV}
\begin{tabular}{c|c|c|c|c|c|c|c}\hline\hline
Classes  & Innermost & Outermost & Low $T$  & High $T$ & DP & $W$ & Typical cases \\ \hline
$W^{1-}$     & unstable & unstable & unstable large & unstable small  & in pairs & $-1$ & Schwarzschild BH \\
$W^{0+}$    & stable     & unstable & unstable large+stable small    & no & one more GP & $0$ & RN BH \\
$W^{0-}$     & unstable & stable     & no & unstable small+stable large  & one more AP & $0$ & Schwarzschild-AdS BH \\
$W^{1+}$    & stable     & stable     & stable small  & stable large & in pairs & $+1$ & RN-AdS BH \\ \hline\hline
\end{tabular}
\end{table*}
\twocolumngrid

\end{document}